\documentclass{article}

\usepackage[english]{babel}
\usepackage{graphicx}
\usepackage{dcolumn}
\usepackage{bm}
\usepackage[latin9]{inputenc}
\usepackage{amsmath}
\usepackage{authblk}
\usepackage{cite}

\begin{document}
		
\title{Procedure for experimental reconstructions of 3D atomic structures from electron microscopy images using atom-counting and a Bayesian genetic algorithm}

\author[1,2]{Annick De Backer}
\affil[1]{EMAT, University of Antwerp, Groenenborgerlaan 171, 2020 Antwerp, Belgium}
\affil[2]{NANOlab Center of Excellence, University of Antwerp, Groenenborgerlaan 171, 2020 Antwerp, Belgium}

\author[1,2,*]{Sandra Van Aert}

\author[3]{Christel Faes}
\affil[3]{I-BioStat, Data Science Institute, Hasselt University, Hasselt, Belgium}

\author[4]{Peter D. Nellist}
\affil[4]{Department of Materials, University of Oxford, Parks Road, OX1 3PH Oxford, United Kingdom}

\author[5,6,*]{Lewys Jones}

\affil[5]{Advanced Microscopy Laboratory, Centre for Research on Adaptive Nanostructures and Nanodevices (CRANN), Dublin 2, Ireland}
\affil[6]{School of Physics, Trinity College Dublin, The University of Dublin, Dublin 2, Ireland}
\affil[*]{\textit{E-mail: sandra.vanaert@uantwerpen.be,lewys.jones@tcd.ie}}

\maketitle


\begin{abstract}
We introduce a Bayesian genetic algorithm for reconstructing atomic models of nanoparticles from a single projection using Z-contrast imaging. The number of atoms in a projected atomic column obtained from annular dark field scanning transmission electron microscopy (ADF STEM) images serves as an input for the initial three-dimensional (3D) model. The novel algorithm minimizes the energy of the structure while utilizing \textit{a priori} information about the finite precision of the atom-counting results and neighbor-mass relations. The results show excellent prospects for obtaining reliable reconstructions of beam-sensitive nanoparticles during dynamical processes from images acquired with sufficiently low incident electron doses.
\end{abstract}


\section*{Introduction}

It is commonly accepted that the three-dimensional (3D) atomic structure of metallic nanoparticles determines their catalytic properties \cite{Narayanan2004, Tao2010, Calle-Vallejo2014, Barron2015,Liu2021}. Indeed, the presence of highly undercoordinated atoms or stepped facets at the surface govern many catalytic reactions. A quantitative characterization of the atomic configuration at the surface is therefore essential to reveal the active sites of the nanoparticle where reactant molecules are preferentially adsorbed, to understand the mechanisms of the catalytic behavior, and to improve the performance of these systems. Atomic resolution annular dark field scanning transmission electron microscopy (ADF STEM) has become an invaluable tool for imaging metallic nanostructures \cite{Fujita2012,Yankovich2014,Jones2014,DeBacker2015a,Altantzis2019}. In this context, electron tomography has been used to provide insights in the 3D shape of nanostructures \cite{VanAert2011,Bals2014a,Miao2016}, but this technique requires a significant electron dose for the multiple projections. Consequently, this approach is not feasible when investigating small beam-sensitive catalysts or dynamical processes. Therefore, an alternative method has been developed where the 3D atomic structure is reconstructed from a single ADF STEM projection \cite{Bals2012,Jones2014,DeBacker2017}. For this purpose, the number of atoms contained in an atomic column along the third dimension is retrieved from an ADF STEM image. These atom counts are used to create an initial atomic model which serves as an input for an energy minimization to obtain a relaxed 3D reconstruction of a nanostructure. The validity of this atom-counting/energy minimization method has qualitatively been verified using electron tomography and is applied to study several systems \cite{DeBacker2017,Altantzis2019}.\\
\begin{figure}[htb]
	\centering
	\includegraphics[width=0.49\textwidth]{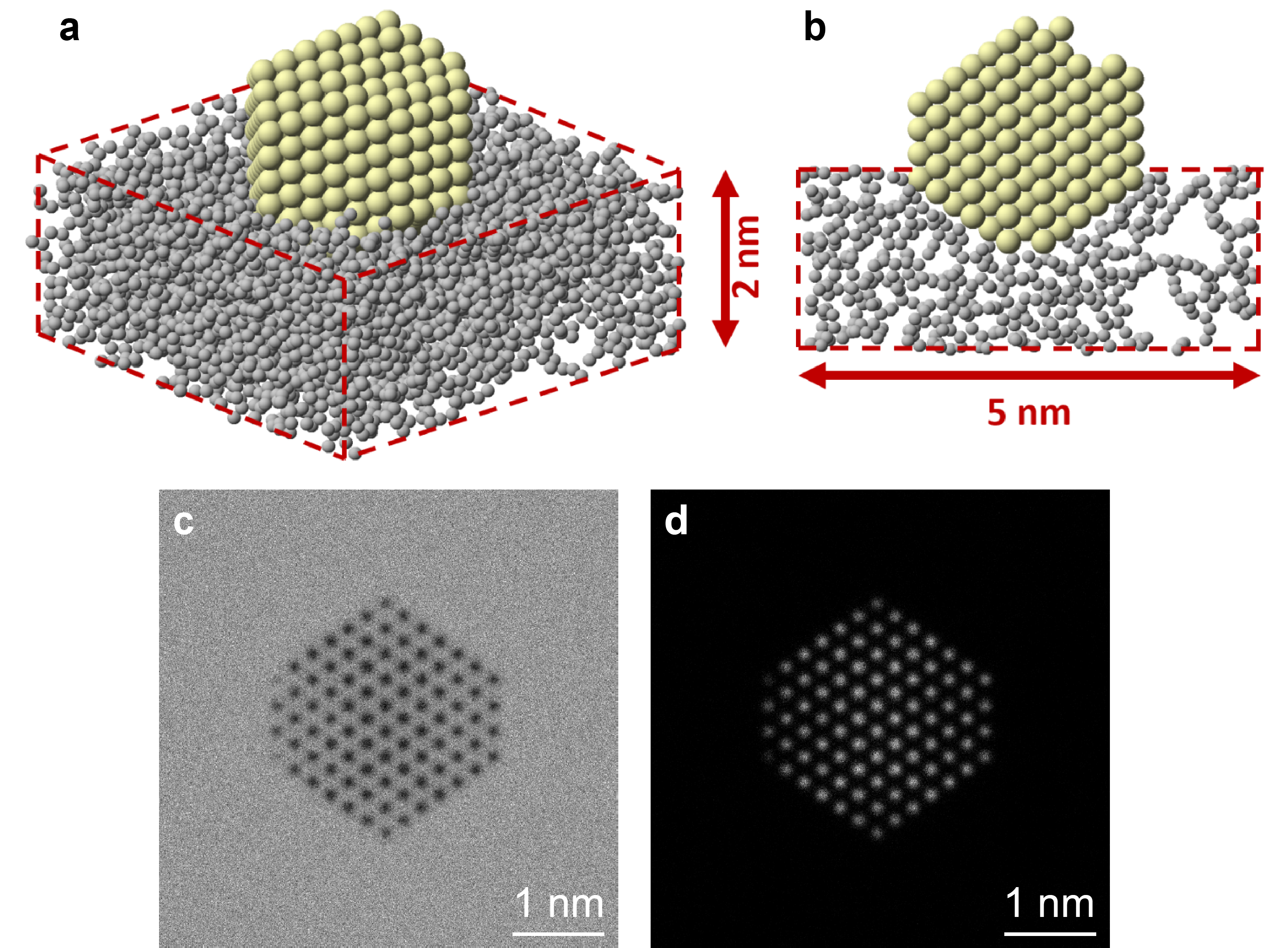}
	\caption{\textbf{Simulated Pt nanoparticle partially embedded in an amorphous carbon support.} \textbf{a} 3D model and \textbf{b} cross-section, \textbf{c} simulated ABF STEM image and \textbf{d} simulated ADF STEM image at $10^4$~e$^-$/\AA$^2$.}
	\label{fig:fig1}
\end{figure}
\begin{figure*}[htb]
	\centering
	\includegraphics[width=\textwidth]{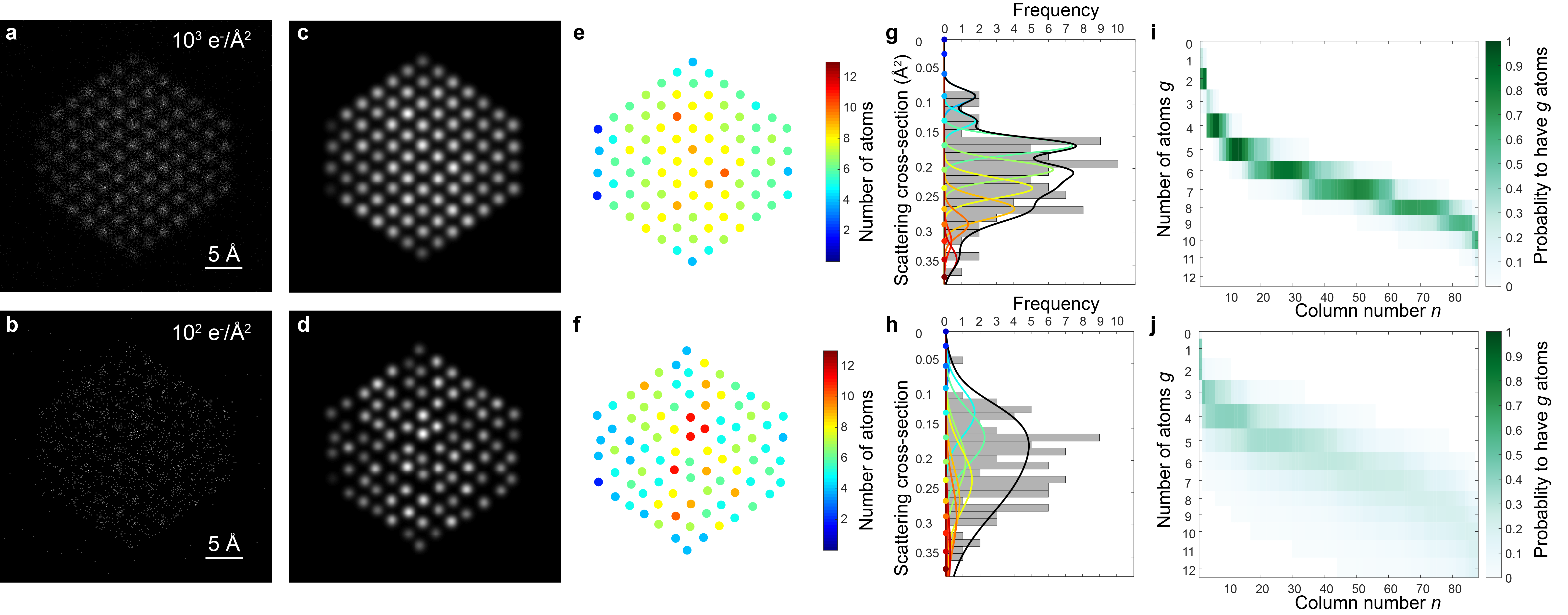}
	\caption{\textbf{Statistical atom-counting analysis and construction of the probability matrix.} \textbf{a-b} Simulated ADF STEM images of the Pt nanoparticle embedded in the carbon support along the [110] zone axis for an incident electron dose of $10^3$~e$^-$/\AA$^2$ and $10^2$~e$^-$/\AA$^2$. \textbf{c-d} Refined parametric models of the images shown in a-b. \textbf{e-f} Estimated number of atoms in each atomic column for the images shown in a-b. \textbf{g-h} Histograms of SCSs for the images shown in a-b. The decomposition into overlapping Gaussian components is shown in color, corresponding to the number of atoms. The black curve shows the estimated mixture model. \textbf{i-j} Probability matrices indicating the probability for a column $n$ to contain a specific number of atoms $g$.}
	\label{fig:fig2}
\end{figure*}
Two possible approaches for the energy minimization are nowadays available. Using the first approach, the energy is minimized by shifting the atomic columns up and down while keeping the number of atoms in a column fixed to the outcome of the atom-counting procedure \cite{Jones2014,Geuchies2016}. The second approach consists of a full molecular dynamics simulation to relax the particle's structure \cite{DeBacker2017,Altantzis2019,Liu2021}. The first method is potentially too restrictive by ignoring the finite atom-counting precision, especially at lower doses. The expected inevitable counting imprecision in this case \cite{DeBacker2015a}, will likely result in slightly more roughness at the reconstructed atomic surface in the direction parallel to the beam direction \cite{Jones2014,DeBacker2017}. On the other hand, the second method runs the risk of ending up in a global energy minimum and violating the physical constraints of the experimental observation. Both approaches hamper a reliable 3D reconstruction of the atomic configuration at the surface, especially for images acquired at lower doses. Here we propose a new method which includes the finite atom-counting precision via a Bayesian inference scheme to improve the 3D atomic models for small nanostructures. Bayesian methods are powerful tools in which a priori information is rationally combined with the observed data and have been successful in other fields of science including business, computer science, economics, educational research, environmental science, epidemiology, genetics, geography, imaging, law, medicine, political science, psychometrics, public policy, sociology, and sports \cite{Gelman2014,Lawson2012, Puri2016,Yamashita2018}. Next to the finite atom-counting precision, the incorporation of additional prior knowledge from neighbor-mass relations will be beneficial when reconstructing atomic models from extremely low dose ADF STEM images. This prior knowledge is fused into a genetic algorithm which uses atom-counting results as an input for reconstructing the 3D atomic structure. In this work, we introduce this advanced Bayesian genetic algorithm. Via an extensive simulation study, the quality of the obtained reconstructions is quantitatively evaluated in terms of the reliability with which the surface atoms can be reconstructed in 3D. In the last part, the algorithm is applied to retrieve 3D atomic models from an experimental time-series of a Pt nanoparticle.\\


\section*{Results and Discussion}

\subsection*{Principle of Bayesian genetic algorithm}

To count the number of atoms, a very well-established approach is utilized \cite{VanAert2011,VanAert2013,DeBacker2013,Jones2014,DeBacker2017}. In the past years, the high accuracy and precision of this method have been demonstrated for nanocrystals of arbitrary shape, size, and atom type. For atom-counting, so-called scattering cross sections (SCSs), corresponding to the total intensity of electrons scattered toward the ADF detector for every atomic column, have been introduced in ADF STEM. These SCSs can be measured using statistical parameter estimation theory \cite{VanAert2009a,DeBacker2016} or by integrating intensities over the probe positions in the vicinity of a single column of atoms \cite{E2013}. For our simulation study, SCSs are determined from noise realizations at different doses of a simulated ADF STEM image of a Pt nanoparticle partially embedded in a carbon support, illustrated in Fig.~\ref{fig:fig1}(a-b). The created particle largely resembles the Wulff construction solid for Pt$_{586}$ (4 atoms along each edge) and was modified to include several diagnostic features of interest commonly observed for catalytic metallic nanoparticles including a surface adatom, a surface vacancy, a terrace edge, a small area of \{110\} face, and rounded corners. With these modifications the particle model contains 587 atoms. The slab of amorphous carbon measures $5~\mathrm{nm}\times 5~\mathrm{nm} \times 2~\mathrm{nm}$ and the geometry follows the work of reference \cite{Ricolleau2013}. Image simulations were performed for the Pt particle viewed along the [110] zone axis. The details of the simulations are described in the Methods Section. Fig.~\ref{fig:fig1}(c) shows the annular bright field (ABF) STEM image, illustrating the presence of the carbon support and Fig.~\ref{fig:fig1}(d) shows the ADF STEM image using an electron dose of $10^4$~e$^-$/\AA$^2$. In Figs.~\ref{fig:fig2}(a-b) simulated ADF STEM images are shown using lower incident electron doses corresponding to $10^3$~e$^-$/\AA$^2$ and $10^2$~e$^-$/\AA$^2$, respectively. The simulated ADF STEM images can be modeled as a superposition of Gaussian functions using the StatSTEM software \cite{DeBacker2017}. The refined models for the simulated ADF STEM images are shown in Figs.~\ref{fig:fig2}(c-d). From the estimated model parameters, the SCSs are determined for each atomic column and are represented in a histogram in Figs.~\ref{fig:fig2}(g-h).
In a subsequent analysis, the distribution of the SCSs of all atomic columns is decomposed into overlapping normal distributions, \textit{i.e.}\ a Gaussian mixture model, as illustrated in Figs.\ \ref{fig:fig2}(g-h) \cite{DeBacker2013,VanAert2013}. The locations of the normal distributions are matched to the expected SCS values from simulations for a column containing $g$ atoms \cite{Dewael2017} and their widths are determined. It should be noted that this procedure only depends on the values of the estimated scattering cross-sections and is independent of the subjective choice of bins in the histogram for visualization. The number of atoms in each projected atomic column in Figs.\ \ref{fig:fig2}(e-f) is then obtained by assigning the SCS to the component that generates this SCS value with the highest probability. More details on the atom-counting methodology are provided in the Methods Section. The width of the normal distributions reflects the finite precision of the atom-counting results and will be used as prior knowledge for reconstructing the 3D atomic structure. As an input we need the probability that an atomic column contains a specific number of atoms which can be defined from the decomposition into normal distributions, illustrated in Figs.\ \ref{fig:fig2}(g-h). Indeed, the normal distribution functions describe the probability that component $g$ generates the $n$th SCS, \textit{i.e.}\ $p(SCS_n|g)$, and by using Bayes' theorem, the probability that the $n$th SCS has $g$ atoms, \textit{i.e.}\ $p(g|SCS_n)$, can be computed:
\begin{align}
p(g|SCS_n) = \frac{p(g)p(SCS_n|g)}{p(SCS_n)} = \frac{p(g)p(SCS_n|g)}{\sum_g p(g)p(SCS_n|g)}.\label{Bayes}
\end{align}
Equal probabilities are assigned to the probability for having $g$ atoms in a column $p(g)$. The value of $p(g)$ depends on the maximum number of $g$ that is considered. However, for the computation of the probability $p(g|SCS_n)$, it is irrelevant since a constant $p(g)$ cancels in Eq.\ (\ref{Bayes}). The probability $p(g|SCS_n)$ is visualized by a probability matrix in Figs.\ \ref{fig:fig2}(i-j). From the representation of the Gaussian mixture model on top of the histogram, the relation between the probability matrix and the width of the normal distributions is clear.\\
For improving the quality of the reconstructions further, we can include even more relevant prior knowledge. For spherical convex nanoparticles, we can include neighbor-mass relations because abrupt discontinuities are highly non-physical \cite{Aarons2017}. The neighbor-mass matrix helps to predict the column mass based on the average mass of the neighboring columns. For small nanoparticles, we propose a diagonal neighbor-mass matrix. The matrix is visualized in Fig.~S2 of the Supplementary Information. The probability profile is Gaussian and the width is chosen such that the interval $\pm 1$ atoms contains $80\%$ of the probability. The normalized neighbor-mass probability, $p(g|NB_n)$ with $NB_n$ indicating the average neighbor-mass, is combined with the probability matrix accounting for the atom-counting reliability $p(g|SCS_n)$, in order to take the two types of prior knowledge into account:
\begin{align}
p\left(g|SCS_n \cap NB_n \right) = \frac{p(g|SCS_n)p(g|NB_n)}{\sum_g p(g|SCS_n)p(g|NB_n)}.\label{priorknowledge}
\end{align}
These probability matrices are used as input of the prior knowledge for the genetic algorithm that we will use to reconstruct the 3D atomic structure of the nanoparticle, hence the name Bayesian genetic algorithm. Genetic algorithms are powerful tools for solving large optimization problems where finding a direct solution is not possible \cite{Goldberg1989,Mitchell1996,Johnston2003,Dugan2009,Rossi2009,Wu2014,Puri2016,Yu2016}. A genetic algorithm is an iterative process where first a population of randomly generated individuals is created. In our algorithm, this initial population is generated by randomly modifying the number of atoms and the height offset of the atomic columns of a 3D starting configuration within a certain mutation range. This 3D starting configuration is obtained by positioning the atoms (\textit{i.e.}\ the outcome of the atom-counting procedure) in each atomic column parallel to the beam direction and symmetrically around a central plane. A population size of 500 is used in all calculations, the count mutation range equals 1, and the height mutation range for the offset of a column equals a lattice step, \textit{i.e.}\ the interatomic spacing between the Pt atoms along the [110] direction. In each iteration, \textit{i.e.}\ a generation, the fitness of every individual in the population is evaluated by the cost-function of the optimization problem. The individuals with the best cost-function values are selected from the current population, and a new complete population of candidate solutions is formed by recombining and mutating the selected individuals. The fraction of the population that is used for the recombination step equals 50\%. For each recombined member, two parents are randomly chosen from the selected individuals, and cross-over is performed by randomly selecting columns from both parents. A mutation density of 2\% is included to avoid ending up immediately in a local minimum by randomly modifying the number of atoms and height offset for 2\% of the atomic columns in each new member. In addition to the usual iterations over many breeding generations, in this work we introduce a second loop to provide for multiple unique starting initializations, specifically to reduce the risk of finding only local-minima solutions. More details of the genetic algorithm are provided in the Methods Section. The cost-function $\chi$ that we use to evaluate the candidate solutions within the different generations of the genetic algorithm is given by:
\begin{align}
\chi = \frac{\sum E_a}{\sum_n g_n}\cdot \left(1+\sqrt[n]{\prod_n p(g_n|\text{prior knowledge})}\right), \label{modelprobability}
\end{align}
where $\sum E_a$ is the sum of the energies per atom given by the EAM potential \cite{Foiles1986,Lee2003} and $\sum_n g_n$ is the total particle mass. This cost-function consists of two factors where the first represents the average energy per atom which we wish to minimize. The second factor takes into account the probability of the candidate solution based on the prior knowledge (Eq.(\ref{Bayes}) or Eq.(\ref{priorknowledge})). This model probability itself is based on the geometric mean of the probabilities of each individual column and needs to be maximized. Since the average energy per atom is negative, this cost-function is minimized. 
\begin{figure}[htb!]
	\centering
	\includegraphics[width=0.49\textwidth]{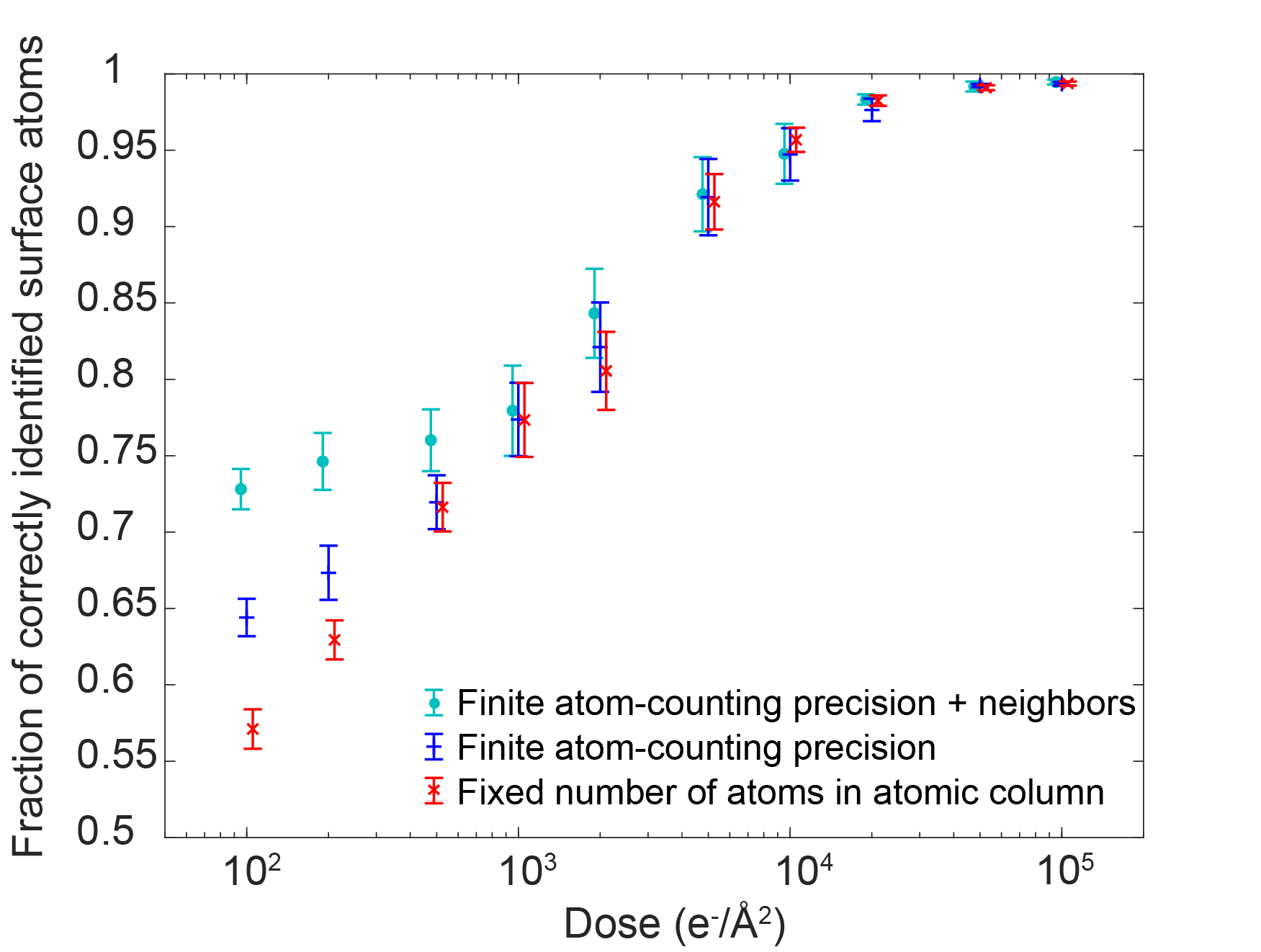}
	\caption{\textbf{Quantification evaluation of the reconstructed 3D models. } Fraction of the surface atoms that are correctly reconstructed in 3D with 95\% error bars when including the finite atom-counting precision and neighbor-mass relations as prior knowledge. As a reference, the results when using a fixed number of atoms in a column are also displayed.}
	\label{fig:fig3}
\end{figure}

\subsection*{Simulation results}
\begin{figure}[htb]
	\centering
	\includegraphics[width=\textwidth]{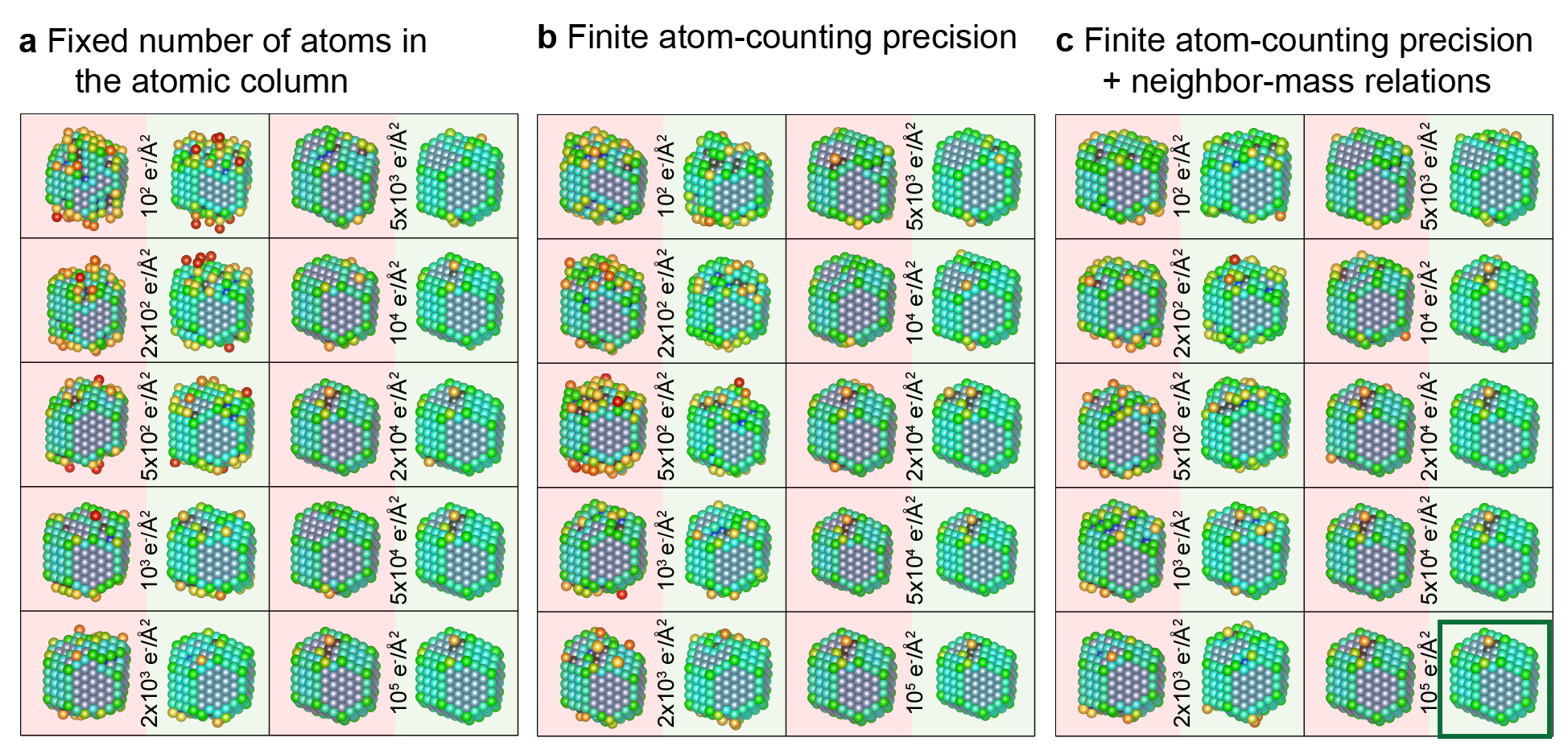}
	\caption{\textbf{Visualization of the reconstructed 3D atomic models.} A lower bound (red background) and upper bound (green background) of a 80\% prediction interval are represented for the reconstructions when using \textbf{a} a fixed number of atoms in the atomic column during the reconstruction, \textbf{b} the finite atom-counting precision, and \textbf{c} the finite atom-counting precision and the neighbor-mass relations. The reconstruction in the box in the lower right corner corresponds to the ground truth model.}
	\label{fig:fig4}
\end{figure}

In order to evaluate the quality of the reconstructions using the Bayesian genetic algorithm, an extensive simulation study is carried out. For this purpose, the electron dose is varied between $10^2$~e$^-$/\AA$^2$ and $10^5$~e$^-$/\AA$^2$ and 30 noise realizations are generated at each electron dose. Fig.~S3 in the Supplementary Information summarizes the atom-counting results obtained following the methodology illustrated in Fig.\ \ref{fig:fig2} and which are used as an input for our Bayesian genetic algorithm. The results of the reconstructions are quantitatively summarized by comparing the reconstructed 3D models with the ground truth model. As a criterion to evaluate the 3D atomic models, we used the fraction of surface atom positions, \textit{i.e.}\ with coordination number less than 12, that are correctly defined in 3D. These are the atoms which are of interest for catalysis. Fig.~\ref{fig:fig3} shows this fraction for the reconstructed atomic models. As a reference, we also included the fraction following the approach where the number of atoms is fixed to the outcome of the atom-counting procedure and where during the reconstruction the atomic columns are only shifted up and down. A significant, vast improvement for the reconstructed surface atoms is observed when including the finite atom-counting precision and the neighbor-mass relations, especially for the lower incident electron doses where we see an improvement from 57\% to 73\%. In Fig.~S4(a) of the Supplementary Information, the results obtained when using the neighbor-mass relations only are also included, next to the results of Fig.~\ref{fig:fig3}. The quality of the neighbors-only reconstructions is dose-independent and significantly lower at the higher incident dose values. In this case, there are no constraints set by the experimental observations and the reconstruction is more determined by a lower energy solution. The high performance at low doses for the neighbors-only reconstructions is partially owing to the choice of a relatively well faceted, low energy particle. These results illustrate that prior knowledge about the finite atom-counting precision is essential in the Bayesian genetic algorithm and that it is the combination of both types of prior knowledge,
that lead to the improved performance as illustrated in
Fig.~\ref{fig:fig3}.\\
In order to evaluate the reconstructed 3D atomic models in a bit more detail, the 4th worst and 4th best reconstructions of the 30 noise realizations at each dose can be visualized as $80\%$ prediction interval for the reconstructions. The 4th worst and 4th best reconstructions are selected based on the percentage of correctly reconstructed surface atoms. These intervals are shown in Fig.~\ref{fig:fig4} and Fig.~S4(b). The colors of the atoms correspond to the coordination number and the reconstruction in the box in Fig.~\ref{fig:fig4}(c) corresponds to the ground truth model. The coordination number serves as a powerful predictor for surface adsorption strength on Pt nanoparticles, and hence as a predictor of chemical activity \cite{Calle-Vallejo2014,Calle-Vallejo2014a,Calle-Vallejo2015,Calle-Vallejo2015a}. Even for the lower doses in Fig.~\ref{fig:fig4}(c), the shape is very well reconstructed and a vast improvement is observed when including relevant prior knowledge resulting in less roughness from the finite atom-counting precision at the surface as compared to the low dose reconstructions in Fig.~\ref{fig:fig4}(a), where the number of atoms in an atomic column has been kept fixed. 

\subsection*{Experimental results}
\begin{figure}[htb]
	\centering
	\includegraphics[width=0.49\textwidth]{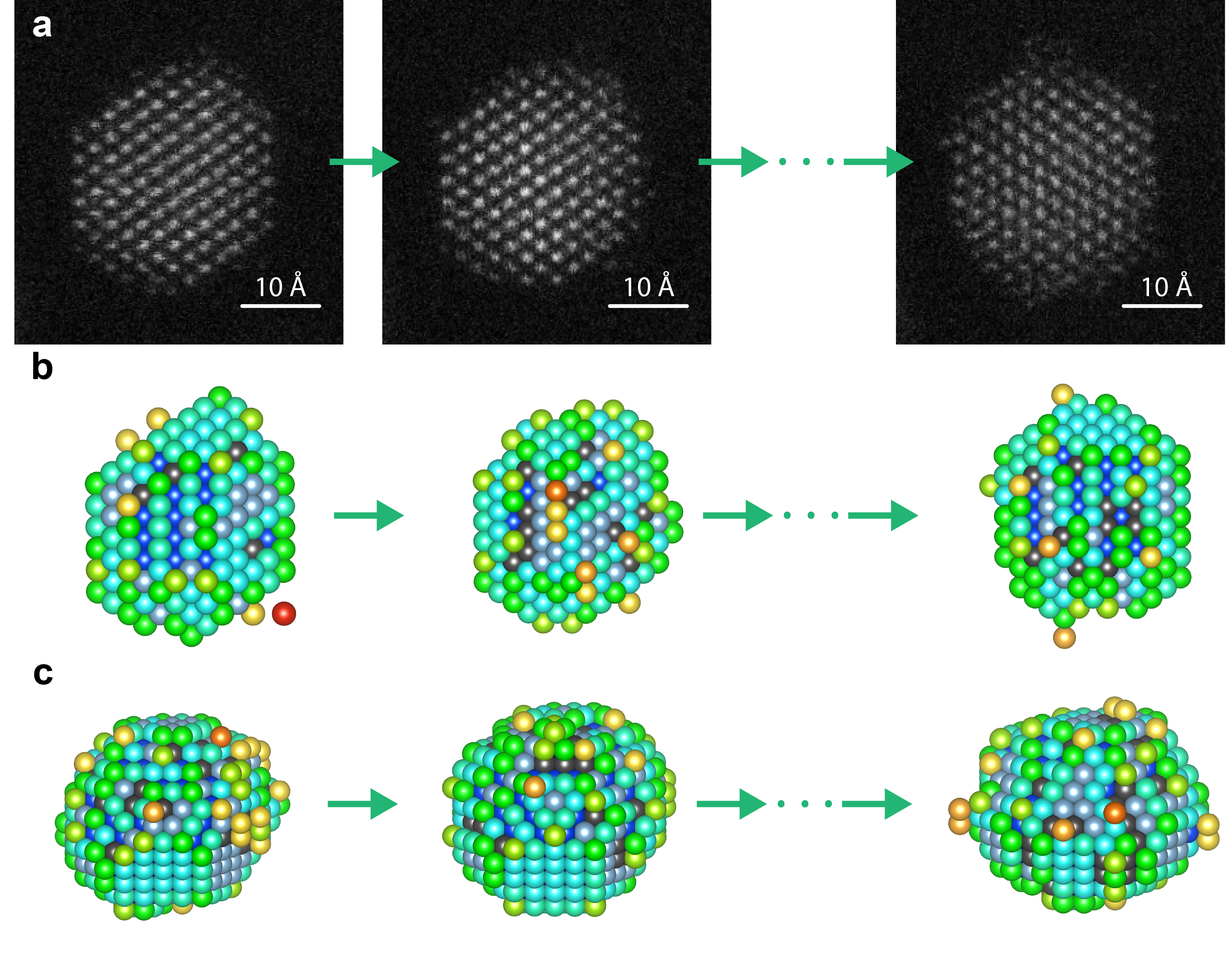}
	\caption{\textbf{Analysis of an experimental time series of a catalyst Pt nanoparticle.} \textbf{a} ADF STEM time series. \textbf{b} Corresponding reconstructed 3D atomic models for the time sequence shown in a viewed along the beam direction. \textbf{c} Rotated models to show the dominant surface facets.}
	\label{fig:fig5}
\end{figure}

As a last part of this work, we apply the Bayesian genetic algorithm to 25 frames of an experimental time series of a catalyst Pt nanoparticle \cite{Aarons2017}. The experimental details and corrections for scan noise and tilt are described in the Methods Section. To reliably count the number of atoms from the time series of images, we used a hidden Markov model which explicitly describes the possibility of structural changes over time \cite{Dewael2020,Dewael2020a}. The atom-counting results from each single frame have been used as an input for our Bayesian genetic algorithm in which we utilize the finite atom-counting precision and neighbor-mass relations. The reconstructed models are schematically represented in Fig.~\ref{fig:fig5}. The ADF STEM images and corresponding reconstructed models for all frames are shown in Supporting Figs.~S5 and S6. This approach enables a reliable 3D quantification of the structural changes of the Pt nanoparticle under the electron beam. From the evaluation of the coordination numbers (Fig.~S7 of the Supplementary Information), we can conclude that although each image has unique noise and that the structure is moving under the electron beam, the number of atoms with the same coordination number is consistent throughout time. Since these coordination numbers are very important to relate to the catalytic properties, it is important to point out here that the small changes clearly do not change the overall catalysis-relevant information that we can extract.

\section*{Conclusions}
To summarize and conclude, we introduced a powerful alternative to the initially developed atom-counting/energy minimization method for the 3D reconstruction of nanoparticles from a single viewing projection. This newly designed Bayesian genetic algorithm takes advantage of the finite atom-counting precision and neighbor-mass relations during the reconstruction. This results in more reliable reconstructions of the 3D atomic structure, especially at lower incident electron doses below $10^4~e^-/$\AA$^2$. The increased quality of the 3D atomic models has been validated by a quantitative evaluation of the reconstructed surface atoms. This result shows great promise to use these reconstructions to predict the adsorption properties of catalytic nanoparticles.

\section*{Methods}
\label{Methods}

\subsection*{STEM Simulations}
Image simulations were performed using the MULTEM package \cite{Lobato2015,Lobato2016}. An acceleration voltage of 200 kV, a semi-convergence angle of 22.48~mrad, and a pixel size of 0.124~\AA\ were chosen and averaging over 30 unique phonon configurations was performed. A source size having a FWHM of 1.0~\AA\ was added to further reflect experiments recorded under the same conditions. The full set of simulation settings are listed in Table~1 of the Supplementary Information.

\subsection*{Atom-counting methodology}

Using the StatSTEM software \cite{DeBacker2016}, a parametric imaging model is fitted to the simulated ADF STEM images. This model consists of a superposition of $N$ Gaussian functions and describes the intensity at the pixel $(k,l)$ at the position $(x_k, y_l)$ of the ADF STEM image:
\begin{align}
f_{kl}(\boldsymbol{\theta}) = \zeta + \sum_{n=1}^N \eta_n \exp \left( - \frac{\left(x_k - \beta_{x_n}\right)^2 + \left(y_l - \beta_{y_n}\right)^2}{2\rho^2}\right)
\end{align}
with $\zeta$ a constant background accounting for the amorphous carbon support, $\rho$ the width of the Gaussian peak, $\eta_n$ the column intensity of the $n$th Gaussian peak, and $\beta_{x_n}$ and $\beta_{y_n}$ the $x$- and $y$-coordinate of the $n$th atomic column, respectively. The unknown parameters of the imaging model are given by the parameter vector:
\begin{align}
\boldsymbol{\theta} = \left(\beta_{x_1},\ldots,\beta_{x_N},\beta_{y_1},\ldots,\beta_{y_N},\rho,\eta_1,\dots,\eta_N,\zeta\right)
\end{align}
and are estimated in the least square sense. From the obtained estimated parameters $\hat{\boldsymbol{\theta}}$, the estimated scattering cross-sections (SCSs) can be calculated from the volumes under the Gaussian peaks above the background \cite{VanAert2009a,DeBacker2016}:
\begin{align}
\mathrm{SCS}_n = 2\pi\hat{\eta}_n\hat{\rho}^2_n.
\end{align}
The estimated SCSs are visualized in a histogram in Fig.\ 2(g,h) and are regarded as independent statistical draws from a so-called Gaussian mixture model. In a sense, the assumption of independent statistical draws implies that cross-talk between neighboring atomic columns is not significantly contributing \cite{Allen2003,Fertig1981,Nellist1999,Martinez2018}. The model is defined as a superposition of Gaussian components and describes the probability that a specific SCS value is observed. The probability density function of a mixture model with $G$ components can parametrically be written as:
\begin{align}
f_\mathrm{mix}\left(\mathrm{SCS}_n;\boldsymbol{\Psi}_G\right) = \sum_{g=1}^G \pi_g \frac{1}{\sqrt{2\pi}\sigma_\mathrm{GMM}}\exp\left(\frac{-\left(\mathrm{SCS}_n-\mu_g\right)^2}{2\sigma_\mathrm{GMM}^2}\right).\label{GMM}
\end{align}
The locations $\mu_g$ of the normal distributions are matched to the expected SCS values for a column containing $g$ atoms obtained from image simulations performed for a Pt crystal in [110] orientation up to 30 atoms thickness. The  settings for the frozen lattice multislice simulation are the same as for the simulation of the Pt nanoparticle embedded in the carbon support (listed in Table 1 of the Supplementary Information). The symbol $\boldsymbol{\Psi}_G$ in Eq.(\ref{GMM}) represents the vector containing all unknown parameters in the mixture model with $G$ components:
\begin{align}
\boldsymbol{\Psi}_G = \left(\pi_1,\ldots,\pi_{G-1},\sigma_\mathrm{GMM}\right).
\end{align}
The parameters $\pi_g$ and $\sigma_\mathrm{GMM}$ denote the mixing proportion of the $g$th component and the width of the components respectively. The mixing proportions $\pi_g$ sum up to unity, therefore the $G$th mixing proportion is omitted in the parameter vector $\boldsymbol{\Psi}_G$. The parameters $\boldsymbol{\Psi}_G$ are estimated from the measured SCSs using the maximum likelihood estimator for a given number of components $G$. Here, $G$ equals the maximum thickness of the image simulations of the Pt crystal, \textit{i.e.}\ 30 atoms, since mixing proportions of components that exceed the maximum thickness of the Pt nanoparticle are estimated zero.\\
In principle, the width $\sigma_\mathrm{GMM}$ reflects the finite atom-counting precision. However, it is known that the width of the Gaussian distributions might be underestimated for lower electron doses \cite{Dewael2017}. In order to counterbalance this underestimation, an effective width $\sigma_\mathrm{eff}$ will be used to describe the width of the normal components in the Gaussian mixture model. For this purpose, $\sigma_\mathrm{GMM}$ is evaluated with respect to the expected dose-dependent width $\sigma_\mathrm{dose}$ as explained in more detail in the Supplementary Information.

\subsection*{Genetic algorithm}

The parameters of the genetic algorithm are specified for the reconstruction of small nanoparticles and balanced with the available RAM/CPU resources (desktop computer). A larger \textit{population size} will capture a more diverse spread of solutions and result in a longer computation time. For small particles, a smaller population size can be used. Here, a population size of 500 particle configurations is used which is significantly larger than the number of atomic columns in the particle ($\approx$ 100 atomic columns). A fraction of this population, \textit{i.e.}\ the \textit{recombination fraction or cross-breeding fraction} will survive from one generation to the next and will be used as `parents' for recombining or cross-breeding new solutions. A too small fraction would reduce the diversity in the solution. A too large fraction on the other hand, reduces the amount of new children in each generation. Therefore, a 50\% fraction is chosen which means that every solution with a quality above average will be preserved and selected for breeding the next generation. If only existing parents are used in the recombination step, there is a risk of ending up in a local minimum. For this reason, after each breeding operation, a small percentage of \textit{mutants} for 2\% of the atomic columns can be introduced to inject some randomness in the atomic models. If these traits are an improvement, they will persist, otherwise they will automatically disappear. A very large degree of mutation is undesirable. In such a case, the benefit from the history of the evolution might be lost. The degree of mutation is expressed in terms of the \textit{height mutation range} and \textit{count mutation range}. If these values are too large, very non-physical solutions are proposed. Here we consider a mutation range of $\pm 1$ atom in mass for a given atomic column and $\pm 1$ position height shift or lattice step. This choice is also justified from both the finite atom-counting precision and the expected finite surface roughness. Important to notice here is that if a mutation is beneficial, then a cumulative evolution is possible. In this manner, a solution of $\pm 2$ or more atoms and/or $\pm 2$ or more height changes can be obtained over different generations. Next to the iterations throughout the different generations, a second loop feeds multiple unique starting initializations to the algorithm to reduce the risk of ending with local-minima solutions. For the simulation study, for each reconstruction, 25 unique random starting populations are used in the algorithm. It should be noted that a structure with a better cost-function might be found when increasing this number. For the experimental time series, we used 100 unique populations initializations throughout the reconstruction procedure.\\

\begin{table}[t]
\centering
\begin{tabular}{ll}
\\ \hline
Parameter & Value\\\hline
Population size & 500\\
Recombination fraction &  50\%\\
Mutation density & 2\%\\
Height mutation range & $\pm 1$ lattice step\\
Count mutation range & $\pm 1$ atom\\\hline
\end{tabular}
\caption{Genetic algorithm parameters.}
\label{tab:tab2}
\end{table}  

\subsection*{Experimental Pt series}

The ADF STEM images were recorded on a JEOL ARM200CF fitted with a probe-aberration corrector using an acceleration voltage of 200~kV, a probe convergence angle of 22.48~mrad, an annular detector ranging from 52-248~mrad, a dwell time of 4~$\mu$s and an incident electron dose of $1.38 \cdot 10^4 e^-/$\AA$^2$ per frame. The images of the time series were corrected for drift and other distortions using non-rigid registration \cite{Jones2015}. During the time-series, the Pt nanoparticle tilts slightly away from zone axis orientation and back, which affects the scattering cross-sections \cite{MacArthur2015}. Therefore, the scattering cross-sections of the individual frames need to be compensated for tilt. This is done by using a linear scaling of the scattering cross-sections of the individual frames \cite{Dewael2017}, assuming that the total number of atoms in the nanoparticle remains constant throughout the time series. This assumption is valid since the threshold energy for sputtering Pt atoms from a convex surface with step sites is 379~keV \cite{Egerton2010}, well above the incident electron energy of 200~keV. We therefore do not expect sputtering of atoms from the surface, only surface diffusion \cite{VanAert2019}.

\section*{Data availability}
The datasets and codes that support the findings of this study are available from the corresponding authors upon reasonable request. 

\section*{Acknowledgments}
This work was supported by the European Research Council (Grant 770887 PICOMETRICS to S. Van Aert and Grant 823717 ESTEEM3). The authors acknowledge financial support from the Research Foundation Flanders (FWO, Belgium) through project fundings (G.0267.18N, G.0502.18N, G.0346.21N) and a postdoctoral grant to A. De Backer. L. Jones acknowledges Science Foundation Ireland (SFI - grant number URF/RI/191637), the Royal Society, and the AMBER Centre. The authors acknowledge Aakash Varambhia for his assistance and expertise with the experimental recording and use of characterization facilities within the David Cockayne Centre for Electron Microscopy, Department of Materials, University of Oxford and in particular the EPSRC (EP/K040375/1 South of England Analytical Electron Microscope)

\section*{Author contributions}
A.D.B., S.V.A., P.D.N., and L.J. conceived the idea. L.J. implemented the genetic algorithm. A.D.B. implemented the Bayesian aspect in the genetic algorithm and performed the analysis. C.F. assisted with the formulation of the Bayesian cost-function. A.D.B. wrote the manuscript. All authors reviewed and commented on the manuscript.

\section*{Competing interests}
The authors declare no competing interests.

\section*{Additional information}
Supplementary information available

\bibliographystyle{unsrt}
\bibliography{mybib}

\end{document}